\documentclass[a4paper,fleqn,usenatbib,letters]{mnras}

\usepackage[T1]{fontenc}
\usepackage{ae,aecompl}

\usepackage{graphicx}	
\usepackage{amsmath}	
\usepackage{amssymb}	

\def\hii {H\,{\sc ii}}
\def\kms{km\,s$^{-1}$}
\title[Periodic flares of methanol and water masers]{Discovery of periodic and alternating flares of the methanol and water masers in G107.298+5.639}

\author[M. Szymczak, M. Olech, P. Wolak, A. Bartkiewicz and M. Gawro\'nski]
{M. Szymczak \thanks{E-mail: msz@astro.umk.pl},
M. Olech,
P. Wolak,
A. Bartkiewicz
and M. Gawro\'nski
\\
Centre for Astronomy, Faculty of Physics, Astronomy and Informatics, Nicolaus Copernicus University,\\
 Grudziadzka 5, 87-100 Torun, Poland \\
}

\date{Accepted 2016 March 16. Received 2016 March 5; in original form 2016 March 2}

\pubyear{2016}

\begin{document}
\label{firstpage}
\pagerange{\pageref{firstpage}--\pageref{lastpage}}
\maketitle

\begin{abstract}
Methanol and water vapour masers are signposts of early stages of high-mass star formation but it is generally thought
that due to different excitation processes they probe distinct parts of stellar environments. Here we present observations 
of the intermediate-mass young stellar object G107.298+5.639, revealing for the first time that 34.4\,d flares of 
the 6.7\,GHz methanol maser emission alternate with flares of individual features of the 22\,GHz water maser. High angular 
resolution data reveal that a few components of both maser species showing periodic behaviour coincide in position and 
velocity and all the periodic water maser components appear in the methanol maser region of size of 360\,au.
The maser flares could be caused by variations in the infrared radiation field induced by cyclic accretion
instabilities in a circumstellar or protobinary disc. The observations do not support either the stellar pulsations
or the seed photon flux variations as the underlying mechanisms of the periodicity in the source.
\end{abstract}




\section{Introduction}
The detection of periodic (29$-$668\,d) variations of the 6.7\,GHz methanol maser emission was an intriguing result 
from monitoring observations of high-mass young stellar objects (HMYSOs; \citealt*{goedhart03, goedhart04}; 
\citealt{goedhart09, goedhart14};\citealt{araya10}; \citealt{szymczak11}; \citealt{fujisawa14}; \citealt{maswanganye15, maswanganye16}; 
\citealt*{szymczak15}). The 6.7\,GHz light curves of 16 periodic sources detected so far (\citealt{maswanganye16}) are quite 
diverse from sinusoidal like to flaring with the relative amplitude of 0.2$-$30 (\citealt{fujisawa14}; \citealt{goedhart14}). 
For the few sources mapped so far, the periodic behaviour is seen in individual components of the spectrum arising from a region 
which is only 20$-$50\,per cent of the overall maser structure (\citealt{sanna15}; \citealt{szymczak15}).

Several models have proposed that the regular and periodic behaviour of the methanol masers are related to variations 
in the dust temperature or in the seed photon flux. The temperature of the dust grains can be modulated by (i) periodic accretion 
of material from the circumbinary disc onto a protostar or accretion disc \citep{araya10}, (ii) rotation of hot 
and dense material of the spiral shock wave in the central gap of the circumbinary accretion disc \citep{parfenov14}
or (iii) pulsation of a HMYSO \citep{inayoshi13}. The flux of seed photons can be periodically enhanced in (i) 
a colliding-wind binary \citep{vanderwalt11} or (ii) an eclipsing binary \citep{maswanganye15}.

Since there is no clear observational evidence to support any of these scenarios we have undertaken a search 
for a HMYSO that shows periodic variations in the 6.7\,GHz methanol and 22\,GHz water maser lines. Theoretical
models suggest that the density and temperature regimes in which these maser pumps operate do not overlap 
(e.g. \citealt*{cragg05}; \citealt{gray15}). However, there is observational evidence for velocity and position 
coincidence of individual features of both masers in a few sources (e.g. \citealt{sanna10}; \citealt{bartkiewicz11}). 
The pumping of 6.7\,GHz methanol masers is dominated by radiative excitation, whereas that of 22\,GHz water 
masers depends on collisions with molecular hydrogen (e.g. \citealt{cragg05}; \citealt*{hollenbach13}). 

In this Letter, we announce the discovery of anti-correlated variations of the 6.7\,GHz and 22\,GHz masers
towards G107.298+5.639 (G107 hereafter). The source is located in the L1204/S140 molecular region which contains 
the dark nebula and two \hii\, regions. It is a deeply embedded object driving a quadrupolar outflow, classified 
as an intermediate-mass protostar of 370L$_{\sun}$ (\citealt{sanchez08, sanchez10}) at the distance of 760\,pc
(\citealt{hirota08}). Recently, a periodic (34.6\,d) flaring of the 6.7 GHz methanol maser has been reported 
(\citealt{fujisawa14}).

\section{Observations}
The observations presented here are part of the ongoing maser monitoring programme of HMYSO with the Torun 32-m 
radio telescope \citep*{szymczak14}. G107 was monitored in the 6.7\,GHz methanol and 22\,GHz water lines from 
2014 July to 2015 December and from 2015 July to 2015 December, respectively. The methanol observations
were usually once per day but increased to approximately eight times per day during flares and decreased 
to once per week during quiescent intervals with two gaps of 8 and 16\,d due to schedule constraints.
The water observations were usually once per day with eight gaps of 4$-$5\,d. The observational setup for the
methanol transition observation was the same as reported in \cite{szymczak14} and resulted in a spectral resolution
of 0.09\,\kms\, with a typical rms noise level of 0.35\,Jy before 2015 May and then 0.22\,Jy. 
The methanol flux density calibration was accurate to less than 10\,per cent. 

The half-power beam width of the 32-m telescope at 22\,GHz is 1.7\,arcmin. We used two 4096-channel correlator parts 
with 8\,MHz bandwidth each, which provided a velocity coverage from $-$46 to 8\,\kms\, with respect to the local standard
of rest and a resolution of 0.05\,\kms. The system temperature was from 50 to 250\,K depending on weather conditions
and elevation and the typical rms noise level was 15$-$20\,mK. The data were calibrated by the chopper wheel method and
the temperature scale of the spectra was $T^{*}_\text{A}$ scale derived with an accuracy of about 30\,per cent. 

The target was observed at 6.7\,GHz with the European VLBI Network (EVN\footnote{The European VLBI Network is a joint facility of independent 
European, African, Asian, and North American radio astronomy institutes. Scientific results from data presented in this publication 
are derived from the following EVN project code: ES076.}) on 2015 March 16 and usable data were obtained with the following antennas: 
Jodrell Bank, Effelsberg, Medicina, Onsala, Torun, Westerbork and Yebes. A phase-referencing scheme was applied using J2223+6249
as phase-calibrator with a switching cycle of 195\,s+105\,s (maser + phase-calibrator). The target was observed for 
a total of 4.9\,h. The bandwidth was set to 4\,MHz yielding 180\,\kms\, velocity coverage. Data were correlated with 0.25\,s 
integration time and 2048 spectral channels yielding a  spectral resolution of 0.09\,\kms. The data reduction followed 
standard procedures for calibration of spectral line observations (e.g. \citealt*{bartkiewicz14}), using the {\sc aips} package. 
The source 3C454.3 was used as a delay, rate, and bandpass calibrator. The phase calibrator was imaged and a flux density of 147\,mJy 
was obtained. Self-calibration was performed on the strongest maser component and the solutions were applied to the spectral line data. 
The images have a typical clean beam at full width half-maximum (FWHM) of 5.78$\times$3.68\,mas (PA = $-$78\fdg2) and an rms noise 
level of 1$-$1.5\,mJy\,beam$^{-1}$ in emission free channels. The  total astrometric uncertainty of the maser components was about 3\,mas
according to a procedure reported in \cite{bartkiewicz09}.
 
\section{Results}

\begin{figure}
\resizebox{\hsize}{!}{\includegraphics[]{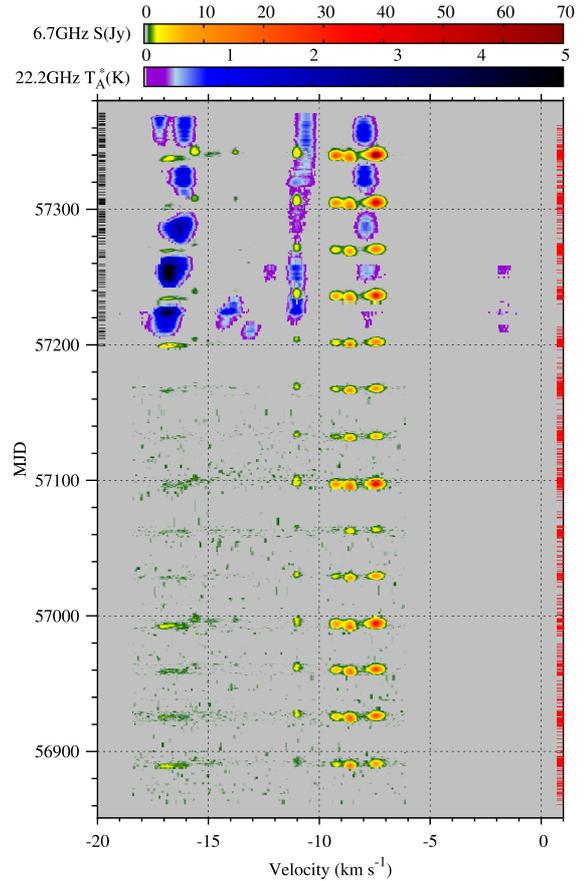}}
    \caption{False-colour image of the 6.7\,GHz methanol maser flux density and the 22\,GHz water maser antenna
             temperature over velocity and time for G107.298+5.639. The velocity scale is relative to the local 
             standard of rest. The horizontal bars in the left (black) and right (red) coordinates correspond to 
             the dates of the observed spectra of water and methanol masers, respectively.}
    \label{fig-dyn_spectra}
\end{figure}

Fig. \ref{fig-dyn_spectra} shows the times series of the 6.7 and 22\,GHz maser spectra.
Four 6.7\,GHz features in the velocity range from $-$11.1 to $-$7.0\,\kms\, were detected in all 14 cycles, 
 whereas a faint ($\leq$3.5\,Jy) emission of 2$-$4 features at velocities from $-$17.2 to $-$13.6\,\kms\, 
was visible only in some cycles. We verified the periodicity using the Lomb-Scargle periodogram (\citealt{scargle82}).
A clear peak at 34.4$\pm$0.8\,d is present in the periodograms of the features
$-$7.4, $-$8.6 and $-$9.2\,\kms. This periodicity is quite consistent with that reported in \cite{fujisawa14}.  
The intensity of the methanol line was extremely variable; the relative amplitude of the strongest feature at $-$7.4\,\kms\, 
reached 120. Here, the relative amplitude is defined as 
($S_{\mathrm{max}} - S_{\mathrm{min}}$)/$S_{\mathrm{min}}$, where $S_{\mathrm{max}}$ and $S_{\mathrm{min}}$ are the 
maximum and minimum flux densities, respectively. The methanol emission was visible over 4$-$12\,d depending on 
the flux density of flare peak.  
The flare curve was almost symmetric, i.e. the ratio of the rise time to the decay time ($R_{\mathrm{rd}}$) was about 
0.93$\pm$0.25 for most features with the exception of the $-$8.6\,\kms\, feature for which $R_{\mathrm{rd}}$=0.59$\pm$0.17.
The average time delays of the maximum flux relative the reference feature of $-$8.6\,\kms\, are 1.7$\pm$0.4, 1.3$\pm$0.3 
and 3.4$\pm$0.6\,d for the features at $-$7.4, $-$9.2 and $-$11.0\,\kms, respectively. This implies a mean separation of 370\,au
to the line of sight and is consistent with that reported by \cite{fujisawa14}.

\begin{figure}
\resizebox{\hsize}{!}{\includegraphics[]{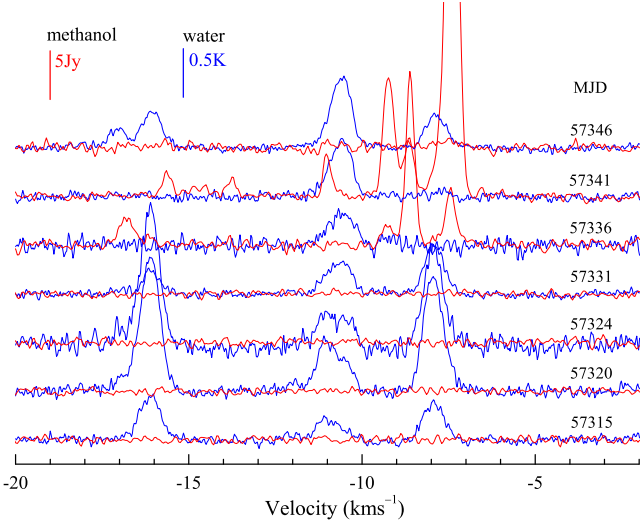}}
    \caption{Selected spectra of 6.7\,GHz methanol maser (red) and 22\,GHz water maser (blue) of G107.298+5.639.
             Red and blue vertical bars show the methanol flux density and the water antenna temperature 
             scales, respectively. The methanol profile taken on MJD 57341 has a peak of 57.7\,Jy at $-$7.43\,\kms\, not shown in the plot.} 
    \label{fig-spectra}
\end{figure}

\begin{figure}
\resizebox{\hsize}{!}{\includegraphics[width=\columnwidth]{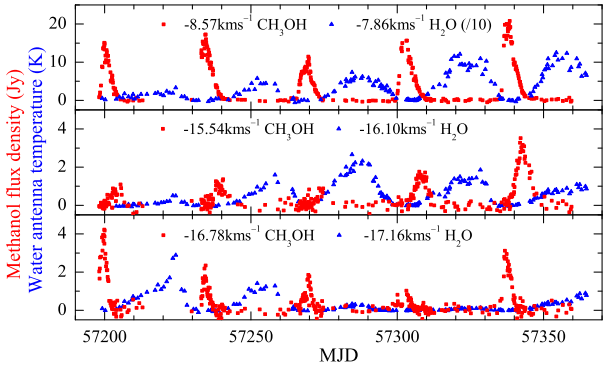}}
    \caption{Light curves of 6.7 and 22\,GHz maser features which coincide in velocity within
             less than 1.1\,\kms.}
    \label{fig-light_curves}
\end{figure}

\begin{table}
 \caption{The parameters of 6.7\,GHz maser clouds. 
\label{clouds}}
\begin{tabular}{c r r r c c}
\hline
Cloud   &  $\Delta$RA &  $\Delta$Dec & $V_{\mathrm{fit}}$ & $FWHM$  & $S_{\mathrm{fit}}$ \\
        &   (mas)     &  (mas)       &  (\kms)            & (\kms)  & (mJy\,b$^{-1}$) \\
\hline  
{\it 1} &   1.122     & -6.729       &  -7.16             & 0.25    &  2.082 \\
{\it 2} &   0.000     & 0.000        &  -7.41             & 0.16    &  7.643 \\
{\it 3} &   0.272     & 0.991        &  -7.46             & 0.59    &  1.642 \\
{\it 4} &   5.223     & -43.979      &  -7.45             & 0.29    &  0.644 \\
{\it 5} &  -15.758    & -9.481       &  -8.09             & 0.24    &  0.103 \\
{\it 6} &  -103.410   & 39.431       &  -8.58             & 0.28    &  2.759 \\
{\it 7} &  -32.649    & -23.026      &  -8.57             & 0.24    &  0.207 \\
{\it 8} &  -46.947    & -38.466      &  -9.02             & 0.26    &  0.370 \\
{\it 9} &  -57.553    & -41.464      &  -9.27             & 0.22    &  0.441 \\
{\it 10}& -392.408    & -94.329      & -10.96             & 0.23    &  0.287 \\
{\it 11}& -334.250    & 240.653      & -16.09             & 0.55    &  0.042 \\
{\it 12}& -324.757    & 283.323      & -16.61             & 0.60    &  0.032 \\
{\it 13}& -262.792    & 321.424      & -16.71             & 0.28    &  0.080 \\              
\hline              
\end{tabular}
\end{table}

\begin{figure}
\includegraphics[width=\columnwidth]{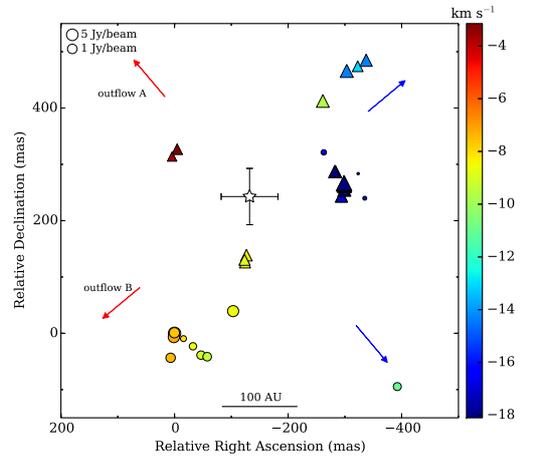}
           \caption{Map of 6.7\,GHz methanol maser components (circles) in G107.298+5.639 obtained with the EVN.
    	    The map origin corresponds to the absolute position RA(J2000) = 22$^{\text h}$21$^{\text m}$26\fs7730, 
    	    Dec.(J2000) = +63\degr51\arcmin37\farcs657 of the strongest maser component. 
    	    The circle size is proportional to the logarithm of maser brightness of component 
    	    and its colour indicates the velocity according to the colour scale on the wedge. 
   	    The triangles indicate the position of the 22GHz water maser components reported by \citet{hirota08}.
    	    The velocity scale of these components is the same as for the methanol line.
    	    A star symbol with the marked positional uncertainty indicates the peak of 1.3\,mm continuum emission 
    	    which is likely the location of an exciting source \citep{palau13}. The red and blue arrows 
    	    refer to the directions of molecular outflows seen at $\sim$20\,arcsec scale \citep{palau11}.}    
    \label{fig-map}
\end{figure}

The water emission ranged from $-$18.3 to $-$1.1\,\kms\, with complex variations in the intensity of all six features.
The emission of the features near $-$16.5 and $-$8.0\,\kms\, significantly dimmed or even disappeared at regular
intervals just when the methanol emission flares peaked at velocities that differ by less than 1.1\,\kms\, 
(Figs \ref{fig-dyn_spectra} and \ref{fig-spectra}). A similar behaviour of the water and methanol emissions was also
seen at a velocity near $-$11.0\,\kms\, before MJD 57290 (Fig. \ref{fig-dyn_spectra}). These water features peaked 
either just in the middle between the methanol maxima or delayed by 2$-$5\,d (Fig. \ref{fig-light_curves}).
It is noteworthy that faint (0.2$-$0.25K) or moderate (0.4$-$0.8K) water maser flares near $-$1.7, $-$12.2, $-$13.0
and $-$13.9\,\kms\,, seen in one or two cycles, peaked only at $\sim$0.5 phase separation from the methanol flares
(Fig. \ref{fig-dyn_spectra}). The relative amplitude of strong water features was up to 40.
The values of $R_{\mathrm{rd}}$ varied in the range of 1.05 to 1.95 from cycle to cycle and no flare curve was observed
with $R_{\mathrm{rd}}<1$.
The strongest water maser feature near $-$16.7\,\kms\, drifted in radial velocity by 0.6\,\kms\, over 2 months 
(Fig. \ref{fig-dyn_spectra}). As the systemic velocity is $-$11.3\,\kms\, (\citealt{sanchez10}) this drift can 
be interpreted as an effect of gas deceleration at the interface between an accelerated outflow and the molecular 
environment of HMYSO.

The results of the EVN observation are summarised in Table \ref{clouds} which lists the following parameters of 
the methanol maser clouds with Gaussian velocity profiles: relative coordinates, peak velocity ($V_{\mathrm{fit}}$), 
line width at half-maximum (FWHM) and  flux density ($S_{\mathrm{fit}}$). The term of cloud is defined as in \cite{bartkiewicz14}.

The 13 methanol maser clouds detected in G107 are mainly grouped in two clusters separated 
by $\sim$400\,mas in the south-east (SE)$-$north-west (NW) direction (Fig. \ref{fig-map}). 
The higher velocity emission ($>-$9.5\,\kms), i.e. redshifted with respect to the systemic velocity
is distributed within the SE cluster (clouds {\it{1$-$9}}) of size 140$\times$70\,mas. The peak brightness temperature 
($T_{\mathrm{b}}$) of maser components in this cluster ranges from $2\times10^8$\,K to $10^{10}$\,K. 
Weak emission (clouds {\it 11$-$13}) at lower velocity ($\le-$15.8\,\kms) from the NW cluster has an arc-like 
morphology of about 70\,mas in extent. Their components have $4\times10^7\le T_{\mathrm{b}}\le10^8$\,K.  
Both maser clusters containing a total of 12 clouds are oriented along an axis of position angle of $-$43\degr\, 
that agrees perfectly with the position angle of the major axis of large-scale ($\sim$20\arcsec) outflow B
traced by CO lines (\citealt{palau11}). 
The isolated methanol cloud {\it 10} near $-$11.0\,\kms\, is located $\sim$360\,mas westward from the SE cluster
and its $T_{\mathrm{b}}$ is $4\times10^8$\,K. It lies very close to the axis of outflow A seen in CO lines
(\citealt{palau11}).

In Fig. \ref{fig-map} the map of methanol maser clouds is superimposed on the 22\,GHz water maser distribution 
observed from 2006 November to 2007 December with VLBI Exploration of Radio Astrometry (VERA) by \cite{hirota08} 
(their tables 1 and 2). The position of the 1.3\,mm continuum emission is showed with error 
bars (\citealt{palau11}) and the directions of molecular outflows (\citealt{palau13}) are marked.
The water masers in G107 are distributed in four clusters from a roughly circular region of size 
$\sim$300\,mas (\citealt{hirota08}). 
The southern cluster of water emission with velocities from $-$8.8 to $-$8.2\,\kms\, lies 95\,mas north to 
$-$8.6\,\kms\, methanol emission (cloud {\it 6} in Table \ref{clouds}). Their intensities vary in opposite way 
(Fig. \ref{fig-light_curves}). 
The water masers in the velocity range from $-$18.1 to $-$16.4\,\kms\, detected in the western cluster 
of arc-like shape are displaced by only 30\,mas from the methanol maser arc-like structure at velocities
from $-$16.7 to $-$16.1\,\kms\, (clouds {\it 11$-$13} in Table \ref{clouds}). We note that the astrometric uncertainties 
of water and methanol maser positions are 0.1\,mas and 3\,mas, respectively, so that part of 
this displacement can be affected by a proper motion because we compare the data sets at two epochs spanned 8\,yr. 
Simultaneous observations of both lines at high angular resolution will be needed to be more definitive.
We preliminary conclude that these water and methanol masers coincide within 20\,au for the assumed distance of 764\,pc
and exhibit alternating flares. The region of co-existence of the two masers is located 140\,pc away from
the central source of infrared emission. 
The water emission at velocities from $-$3.6 to $-$3.2\,\kms\, from the western compact ($\sol$30\,mas) 
cluster was not detected during our monitoring. Three features of water maser at velocities from $-$14.3 
to $-$9.7\,\kms, which appeared in two cycles may come from the NW elongated ($\sim$120\,mas) cluster 
imaged by \cite{hirota08}.

\section{Discussion}
Our monitoring observations provided the first evidence that G107 alternates cyclically between the 6.7\,GHz methanol
maser emission state and the 22\,GHz water maser state with a period of 34.4\,d. The detectable methanol emission
lasts only about 20\, per cent of the period when the water emission drops below the detection level.
The periodic features of both maser lines of G107 coincide within less that 1.1\,\kms\, and at least some clouds of 
both maser species are in the same volume of molecular gas of size 30$-$80\,au. Similarly,  the co-existence of features
of both masers in regions of size 100$-$300\,au is seen for a few objects  (e.g. \citealt{sanna10}, a few spots of 
cluster C in their fig. 3; \citealt{bartkiewicz11}) albeit many interferometric observations demonstrated 
that the methanol and water masers arise from different parts of the HMYSO environments.

It is generally accepted that the 22\,GHz water transition is collisionally pumped in the cooling post-shock gas 
(\citealt*{elitzur89}; \citealt{hollenbach13}), while the 6.7\,GHz methanol transition is pumped by mid-infrared photons
(\citealt{cragg05}). The 22\,GHz transition is inverted over a broad range of physical conditions with a kinetic temperature 
($T_{\mathrm{k}}$) from 200 to 2000\,K, number density of $\le10^8$ to $10^{10}$\,cm$^{-3}$ and fractional abundance higher 
than $10^{-5}$ (e.g. \citealt*{yates97}; \citealt{hollenbach13}; \citealt{gray15}). The 22\,GHz maser gain weakly depends 
on dust temperature ($T_{\mathrm{d}}$) lower than 100\,K, but it is reduced by an order of magnitude for 
$T_{\mathrm{d}}$ = 300K and the pumping is most efficient when $T_{\mathrm{d}}<$T$_{\mathrm{k}}$ (\citealt{yates97}).
Recently published models confirm this trend; the lowest inversion of the 22\,GHz transition occurs for $T_{\mathrm{d}}\approx$ 500K
(\citealt{gray15}). 

The 6.7\,GHz maser transition is radiatively pumped over a broad range of the number density ($10^4 - 10^8$\,cm$^{-3}$) at methanol
fractional abundance greater than $10^{-7}$. The brightness temperature increases rapidly for $T_{\mathrm{d}}$ about 120\,K and then is
nearly constant up to $T_{\mathrm{d}}$=500\,K  (\citealt{cragg05}). It slightly drops with the gas temperature increase 
from 30 to 250\,K.  The 6.7\,GHz emission diminishes when $T_{\mathrm{k}}$ approaches or exceeds $T_{\mathrm{d}}$. 

It is apparent from the above modelled ranges of the physical parameters that the 6.7\,GHz methanol and 22\,GHz water masers
can co-exist in the same regions of G107 for a narrow range of gas density of $1-4\times10^8$\,cm$^{-3}$ and kinetic temperature
of $200-250$\,K. Note that this upper limit of $T_{\mathrm{k}}$ is poorly set as higher values were not investigated. However,
one expects that for higher $T_{\mathrm{k}}$,  the rotational levels are quickly thermalized and the methanol maser is quenched.
No mechanism is known which produces changes of gas density and methanol abundance by an order of magnitude over a region of size 300\,au
on time-scales of 1 week. It seems that only the dust temperature can react sufficiently fast to the variations of the central
object radiation and controls the methanol and water maser intensities. Since the pumping processes are competitive, a pulse of 
the strong infrared radiation quenches collisionally pumped 22\,GHz water transition and enhances the radiatively pumped 6.7\,GHz 
methanol transition. 

Previous observations of G107 (\citealt{sanchez08,sanchez10}; \citealt{palau11,palau13}) suggest that the masers are excited by 
a deeply embedded young stellar object of about 4$-$5M$_{\sun}$ and a bolometric luminosity of 370L$_{\sun}$. This newly born 
star drives a quadrupolar outflow of dynamical age of about 1200\,yr and a faint ultracompact \hii\, region ($<$0.01\,pc) with
characteristics typical for a Class 0 protostar of spectral type B3. The emission of complex molecules probably traces a rotating
disc of size $\sim$300\,au. The gas temperature derived from the thermal methanol lines is about 100$-$150\,K.

\cite{inayoshi13} have proposed that the observed maser flares follow the luminosity variations of HMYSOs which are pulsationally 
unstable when growing under rapid mass accretion with rates $\ga10^{-3}$M$_{\sun}$\,yr$^{-1}$. For the observed period of 34.4\,d
their models give a luminosity 40 times higher than that inferred from observations of G107 (\citealt{sanchez10}). 
Thus we conclude that this mechanism is an unlikely explanation for the periodic maser behaviour.

There is a growing number of embedded Class 0 protostars only detectable at infrared wavelengths in which the cyclic variability 
(\citealt{hodapp15}; \citealt{safron15}) can be related to accretion instabilities developed by interactions between 
the stellar magnetosphere and the accretion disc (\citealt{dangelo12}). In addition to the episodic changes in luminosity on time-scales
of a few years, this mechanism could invoke short period ($\sim$30\,d) modulations with amplitude of 1$-$2\,mag. 
For instance an embedded outflow source IRS\,7 in the L1634 molecular cloud shows periodic flares with a period of $\sim$37\,d and 
amplitude of about 2 mag in the $K_s$ band (\citealt{safron15}). It spends about half the period in a quiescent state that is
comparable to that observed in the archetypal periodic source G9.62+0.20E (\citealt{vanderwalt16}) but is much shorter than in G107.
Nevertheless this scenario can be viable because in addition to a well defined periodicity in G107 we observed 
significant modulations of the maser emission peaks which can be induced by accretion rate variations on longer time-scales.

\cite{araya10} have postulated that the maser gain follows the periodic accretion instabilities induced in a protobinary system.
This mechanism of binary orbit triggered accretion rate (\citealt{artymowicz96}) modelled for the 6.7\,GHz maser line by \cite{parfenov14} 
predicts that the illumination of the disc by the bow shock hot material led to the variation of the dust temperature in the disc 
and regulates the maser brightness with the orbital phase. The detailed results of this modelling are questioned (\citealt{vanderwalt16}) but 
this scenario may be an alternative explanation of the maser behaviour in G107. Using the theoretical values from \cite{artymowicz96} 
we found the observed 6.7\,GHz flare curve similar to the accretion rate curve for a 4 and 3M$_{\sun}$ binary with eccentricity 
of 0.6, a period of 0.09\,yr and a semi-major axis of 0.4\,au. 
Therefore, it seems that the orbital characteristics of a hypothetical binary system in G107 may be similar to those suggested
for short (29.5\,d) period source G12.89+0.49 (\citealt{goedhart09}) and very different from longer ($>$200\,d) period masers such as G9.62+0.20E
(\citealt*{vanderwalt09}). High angular resolution data of G107 (e.g. \citealt{sanchez10}; \citealt{palau11})
do not provide evidence for such a binary (\citealt{sanchez10}; \citealt{palau11}).

Our observations do not support the  hypothesis that the maser flares in G107 are due to variations in the free-free background seed photon flux
produced by a colliding-wind binary (\citealt{vanderwalt11, vanderwalt16}) or an eclipsing binary (\citealt{maswanganye15}). Changes in the
continuum emission of the ultracompact \hii\, with a spectral index of $+$0.5 (\citealt{sanchez08}) should result in correlated variations
of the water and methanol maser intensities.

\section*{Acknowledgements}
We thank the Torun CfA staff and the students for assistance with the observations.
We are grateful to Eric Gerard for reading and useful comments on the manuscript.
This research has made use of the SIMBAD data base, operated at CDS (Strasbourg, France)
and NASA's Astrophysics Data System Bibliographic Services.








\bsp	
\label{lastpage}
\end{document}